\documentstyle[epsf]{mn}

\newcommand{\cidfig}[6]{
    \protect\centerline{
    \epsfxsize=#1\epsffile[#2 #3 #4 #5]{#6}
    }}

\def\et{et\thinspace al.\ }                                 
 
\def\arcsec{$^{\prime\prime}$}

\title[Spectral Synthesis of the Nuclear Region of Seyfert 2 and Radio Galaxies]
      {Spectral Synthesis of the Nuclear Region of Seyfert 2 and Radio Galaxies}

\author[H. R. Schmitt, T. Storchi-Bergmann, R. Cid Fernandes]
       {Henrique R. Schmitt$^{1,2\mbox{$\dagger$}}$\
        Thaisa Storchi Bergmann$^{2\mbox{$\dagger$}}$\thanks{
Visiting Astronomer at the Cerro Tololo
Interamerican Observatory, operated by the Association of
Universities for Research in Astronomy, Inc. under contract with the
National Science Foundation.}\
        and
        Roberto Cid Fernandes Jr.$^3$\thanks{E-mail: 
          schmitt@stsci.edu (HRS); 
          thaisa@if.ufrgs.br (TSB); cid@fsc.ufsc.br (RCF).} \\
        $^1$ Space Telescope Science Institute, 3700 San Martin Drive,
Baltimore, MD21218, USA \\
        $^2$ Instituto de F\'{\i}sica - UFRGS, Caixa Postal: 15051.
CEP: 91501-970 Porto Alegre, RS, Brazil.\\
        $^3$ Departamento de F\'{\i}sica, CFM - UFSC, Campus
Universit\'ario - Trindade, Caixa Postal: 476. CEP: 88040-900
Florian\'opolis, SC, Brazil.}

\begin{document}
\maketitle

\begin{abstract}

We present the results of an optical spectral synthesis analysis for
the nuclei of 20 Seyfert 2 and 4 Radio Galaxies, using a base of stellar
population templates of different ages and metallicities and a
power-law continuum. Compared with the stellar population of elliptical
galaxies, we find that Seyfert 2s usually have a smaller contribution
from old metal rich stars (10 Gyr, Z$\geq$Z$_{\odot}$), and a larger
contribution from stars with 100 Myr.
We also find that the contributions from stars with age $\le
10$\ Myr and from a power law continuum are small, rarely exceeding 5
per cent. These results show that the general assumption of elliptical
galaxies as stellar population templates for these objects is
incorrect, also implying that the excess blue continuum frequently
found in their nuclear spectra is probably due to this template
mismatch. We find a considerable contribution from 100 Myr stars
($\approx 5$ per cent), which can be interpreted from the point of view
of models where the fueling of the AGN is done by
interactions/mergers.

\end{abstract}


\section{Introduction}

From the point of view of spectral studies of active galactic nuclei
(AGN), starlight is an unwanted but unavoidable pollution. Removing
the stellar component is the most critical and uncertain step when
analyzing optical spectra of AGN, since it strongly affects the
residual ``pure'' nuclear spectrum, particularly in the continuum.
This would be enough a reason to study the impact of different
starlight evaluation techniques. Another reason why this step
deserves further attention is that the starlight ``pollution''
contains valuable information on the stellar population of the host
galaxy, which might reveal links between the nuclear activity
phenomenum and the star-formation history in the nuclear regions of
AGN.

In this Paper we explore both these issues by using an adaptation
of stellar population synthesis techniques, so successfully applied
to the study of normal galaxies (e.g., Bica 1988; Schmidt, Bica \& Alloin
1990; Jablonka, Alloin \& Bica 1990, 1992; Bonatto et al. 1998),
to determine the stellar population and the contribution from a
featureless continuum (FC) to the nuclear spectrum of AGN's.
Our main motivations were the intriguing results of Cid
Fernandes, Storchi-Bergmann \& Schmitt (1998, hereafter Paper~I),
where we measured equivalent widths (Ws) of several absorption lines
and continuum fluxes from long-slit spectra of a sample of 42
Seyferts, LINERs, normal and radio galaxies (nearly 500 spectra in
total). The analysis of the radial variation of the Ws in Seyfert 2s
and Radio galaxies showed surprisingly little or no signs of
dilution in the nucleus, implying an FC
contribution to the nuclear spectrum of less than 10 per cent in the
optical range. Later, Storchi Bergmann, Cid Fernandes \& Schmitt
(1998 --- Paper~II) presented a detailed spectral decomposition of
four individual Seyfert 2s, confirming that the contribution from a
nuclear FC is of the order of 10 per cent or less. These results constrast
with the much larger FC fractions (20--50 per cent) usually found in the
literature. The origin of this discrepancy traces back to the fact
that most authors use an elliptical galaxy spectrum as a template
for the nuclear stellar population, a method pioneered by Kosky
(1978). It is thus crucial to establish how appropriate this
procedure is, since it is still widely employed in AGN studies.

FC strengths play an important role in several aspects of AGN
research (line--continuum correlations, polarization, etc). As an
example we mention the current controversy over the origin of the so
called ``second featureless continuum'' component (FC2) in Seyfert
2s (Miller 1994; Cid Fernandes \& Terlevich 1995; Heckman \et 1995, 1997;
Tran 1995b). Indeed, judging from the results of Storchi Bergmann \et
(1998), the whole issue of FC2 might rest upon a question of
alternative FC determination methods, which are ultimately related
to different choices of stellar population templates.

In this Paper we present results from a spectral synthesis of the
nuclear region of 20 Seyfert 2s and 4 Radio Galaxies, using  
spectra in the wavelength range $\lambda\lambda$3500-7000\AA\ and a
synthesis code to quantify the contribution of stars of different
ages and metallicities and an FC to the nuclear spectrum of these
objects.

\section{The Data}

The data used in this Paper are described in Paper~I. They consist of
the nuclear spectra---extracted using an aperture of
2\arcsec$\times$2\arcsec---from long-slit observations of 20 Seyfert 2s
and 4 Narrow Line Radio Galaxies, collected at the 4m telescope at the
Cerro Tololo Interamerican Observatory. The galaxies are:  CGCG420-015,
ESO362-G8, ESO417-G6, Fairall316, IC1816, IRAS11215-2806,
MCG-05-27-013, Mrk1210, Mrk348, Mrk573, Mrk607, NGC1358, NGC1386,
NGC3081, NGC5135, NGC5643, NGC6300, NGC6890, NGC7130, NGC7582,
PKS0349-27, PKS0634-20, PKS0745-19 and 3C33 (the last 4 are the radio
galaxies). As a comparison sample we use the spectrum of the elliptical
galaxy IC4889, observed by us with the same setup as that used for the
other galaxies, and the elliptical galaxy templates E1 and E2 from Bica
(1988).

As input for the synthesis we use the Ws of the absorption lines
CaII~K~$\lambda$3933, CN~$\lambda$4200, G~band~$\lambda$4301 and
MgI~$\lambda$5175, as well as the continuum ratios
$\lambda3660/\lambda5870$, $\lambda4020/\lambda5870$,
$\lambda4510/\lambda5870$ and $\lambda6630/\lambda5870$, measured in
Paper I according to the methodology employed by Bica \& Alloin
(1986a,b 1987) in their study of the stellar populations of normal
galaxies. Paper~I also presented W and continuum ratio measurements for
a number of extranuclear spectra (up to 5--100\arcsec\ away from the
nucleus, depending on the case). A full analysis of the spatial
variations of the stellar population mixtures will be published
elsewhere.

\section{The Spectral Synthesis Code}

For the spectral synthesis we use Bica's (1988) code, modified by
Schmitt, Bica \& Pastoriza (1996) to include continuum ratios and
internal reddening in the search for solutions. The code uses a grid of
Ws and continuum ratios, generated from spectral
distributions of star clusters of different ages and metallicities,
representing the 12 principal components of the age
$\times$\ metallicity plane, defined by Schmidt \et (1991).  These
components are listed in Table 1, to which we added a 13$^{th}$
component: a $F_{\nu}\propto\nu^{-1.5}$\ power law, representing a
canonical AGN continuum (e.g., Kinney \et 1991).

Compared with other technics of stellar population synthesis, which use
libraries of stellar spectra, the method introduced by Bica (1988) has
the advantage of reducing the number of unknown variables. This is due
to the fact that the spectra of star clusters already have implicit
parameters such as the initial mass function and surface gravity of the
stars, eliminating the necessity of their determination and reducing
the synthesis problem to the Age$\times$Metallicity plane.

\begin{table*}
\begin{centering}
\begin{tabular}{llllll}
\multicolumn{5}{c}{Base Elements Used}\\ \hline
HII&  10 Myr & 100 Myr & 1 Gyr & 10 Gyr & Log(Z/Z$_{\odot}$)\\ \hline
   &   X   &   X   &  X  &   X     &  0.6\\
 X &   X   &   X   &  X  &   X     &  0.0\\
   &       &       &  X  &   X     & -1.0\\
   &       &       &     &   X     & -2.0\\ \hline
\end{tabular}
\caption{This table does not include the 13$^{th}$ component, FC of the form
F$_{\nu}\propto\nu^{-1.5}$}
\end{centering}
\end{table*}

The synthesis code works by adding different proportions of the base
elements according to a step, 5 per cent in our case, searching for all
possible combinations of the base elements. The Ws and continuum ratios
of all these combinations are then compared with the observed values. A
combination is considered a solution only when all observed and
synthesized values agree to within a given acceptance window. The
adopted windows were of the order of 1--2 \AA\ for the Ws and 5--10 per
cent for the continuum ratios, similar to our measurement errors. Since
the continuum flux ratios depend on the internal reddening, the
observed values were dereddened by different E(B-V) values in the
0--0.5 range, with steps of 0.02, and compared with the synthesized
values.

Due to the fact that spectral synthesis is a degenerate problem, more
than one acceptable solution can represent the observed data.  We
assume that the final solution is given by the weighted average of all
solutions, for all values of E(B-V) tested. The individual solutions
are weighted by exp$^{-\chi^2/2}$, where
$\chi^2=\sum_{i=1}^{4}(W_{oi}-W_{si})^2/\sigma_{wi}^2 +
\sum_{j=1}^{4}(C_{oj}-C_{sj})^2/\sigma_{cj}^2)$ (W$_{oi}$ and W$_{si}$
are the observed and synthesized Ws, C$_{oj}$ and C$_{sj}$ are the
observed and synthesized continuum ratios, while $\sigma_{wi}$ and
$\sigma_{cj}$ are the measured errors of the Ws and continuum ratios).
The typical ratio of solutions:combinations found is
$4\times10^4$:$6\times10^9$.

It should also be noticed that there is a certain ambiguity between the
HII region and FC base elements, since the HII region can be
approximately represented by a power law F$_{\nu}\propto\nu^0$.
Objects with a $\nu^{-1}$\ FC, for instance, will have their FC
strength spread into these two bins. This is not a problem as long as
care is taken when interpreting the results. The added contributions of
these two components is always an upper limit to the strength of the
FC, or of the HII component.

\section{Results}

In Figure 1 we show the results of the synthesis. The Figure is
separated into panels showing histograms of the percentage light
contribution at $\lambda$5870\AA\ (left) and $\lambda$3660\AA\ (right)
from the different components to the nuclear spectra of the galaxies.
The 13 base elements of the synthesis have been grouped into 8
representative components, which are, from top to bottom: (1) stars
with 10 Gyr and metallicity Z/Z$_{\odot}=0.6$; (2) stars with 10 Gyr
and Z/Z$_{\odot}= 0$; (3) the sum of stars with 10 Gyr and low
metallicity (Z/Z$_{\odot}=-1$\ and $-2$); (4) the sum of the three
components with 1 Gyr; (5) the sum of the two components with 100 Myr;
(6) the sum of the two components with 10 Myr; (7) HII Regions and (8)
the FC. These results, normalized to the flux at $\lambda$5870\AA,
plus the average E(B-V) are also presented in Table 2.

\begin{table*}
\begin{centering}
\begin{tabular}{lrrrrrrrrr}
\multicolumn{5}{c}{Synthesis Results}\\
\hline
Name& 10Gyr& 10Gyr& 10Gyr& 1Gyr & 100Myr & 10Myr &  HII &FC  &E(B-V)\\
& Z/Z$_{\odot}$=0.6& Z/Z$_{\odot}$=0.0& Z/Z$_{\odot}<0$& & & & & &\\
\hline
NGC1358       & 73 & 1  &  0  & 26  &  0  &  0  &  0  &  0  & 0.14 \\
MGC1386       & 56 & 15 &  7  & 12  &  8  &  0  &  0  &  0  & 0.33 \\
NGC3081       & 53 & 13 &  7  & 21  &  2  &  4  &  0  &  0  & 0.14 \\
NGC5135       &  2 & 2  &  9  &  4  & 46  &  6  &  17 &  14 & 0.42 \\
NGC5643       &  6 & 6  &  6  & 47  & 35  &  0  &  0  &  0  & 0.51 \\
NGC6300       & 33 & 28 & 26  & 11  &  2  &  0  &  0  &  0  & 0.43 \\
NGC6890       & 13 & 28 & 14  & 37  &  0  &  6  &  0  &  2  & 0.35 \\
NGC7130       &  5 & 13 & 20  &  3  &  9  &  2  &  44 &  4  & 0.24 \\
NGC7582       & 10 & 9  & 16  & 12  & 42  &  0  &  6  &  5  & 0.60 \\
Mrk348        & 19 & 43 & 20  & 13  &  0  &  2  &  0  &  3  & 0.07 \\
Mrk573        & 70 & 8  &  4  & 13  &  4  &  1  &  0  &  0  & 0.01 \\
Mrk607        & 62 & 10 &  5  & 17  &  6  &  0  &  0  &  0  & 0.14 \\
Mrk1210       & 23 & 24 &  7  & 26  &  0  &  5  &  6  &  10 & 0.09 \\
CGCG420-015   & 50 & 17 & 10  & 19  &  4  &  0  &  0  &  0  & 0.33 \\
IC1816        & 21 & 26 & 17  & 30  &  4  &  0  &  0  &  2  & 0.19 \\
IRAS11215-2806& 27 & 37 & 20  & 12  &  4  &  0  &  0  &  0  & 0.12 \\
MCG-05-27-013 & 56 & 18 & 13  & 13  &  0  &  0  &  0  &  0  & 0.27 \\
Fairall316    & 88 & 2  &  0  &  7  &  3  &  0  &  0  &  0  & 0.17 \\
ESO417-G6     & 39 & 23 &  8  & 27  &  3  &  0  &  0  &  0  & 0.13 \\
ESO362-G8     & 25 & 3  &  0  & 21  & 51  &  0  &  0  &  0  & 0.47 \\
3C33          & 81 & 4  &  3  &  5  &  7  &  0  &  0  &  0  & 0.14 \\
P0349-27      & 75 & 11 &  5  &  7  &  2  &  0  &  0  &  0  & 0.04 \\
P0634-20      & 66 & 5  &  2  & 25  &  2  &  0  &  0  &  0  & 0.34 \\
P0745-19      & 46 & 7  & 12  &  3  & 18  &  0  &  5  &  9  & 0.36 \\\hline
\end{tabular}
\caption{The contributions are normalized to the flux at $\lambda$5870\AA.
The last 4 rows are the Radio Galaxies.}
\end{centering}
\end{table*}

Figure 1 shows how remarkably {\it varied} are the stellar populations
of  Seyfert 2s. Considering only the contribution to the light at
$\lambda$5870\AA\ (left panel), the percentage contribution from stars
with 10 Gyr and Z/Z$_{\odot}=0.6$\ can be as small as 5 per cent or
larger than 60 per cent.  Stars with 10 Gyr and other metallicities, as
well as stars with 1 Gyr or 100 Myr, also have a large range of
percentage contributions to the light at $\lambda$5870\AA, from as
little as 5 to nearly 50 per cent, depending on the object.
The contribution from stars with 10 Myr is $<1$ per cent for 13 galaxies,
while that of HII regions is $<1$ per cent for 16 galaxies. For the
remaining galaxies, the contribution from 10Myr and HII regions is larger.
Among these 7 galaxies we have NGC5135 and NGC7130, which are known
from previous works to present
circumnuclear star-formation (e.g.\ Thuan 1984; Gonz\'alez-Delgado et
al. 1998), as well as Mrk1210, which also shows signatures of a young
population, including a Wolf-Rayet feature in the nuclear spectrum
(Paper~II). The contribution from 10 Myr stars and HII regions to the
spectra of these galaxies is, respectively, 6 and 17 per cent  for NGC5135,
2 and 44 per cent for NGC7130, 5 and 6 per cent for Mrk1210. The remaining
galaxies with $>$1 per cent contribution from 10 Myr stars are NGC3081,
NGC6890, Mrk348 and Mrk573, which have 5, 6, 2 and 1 per cent contribution,
respectively. In the case of HII regions, only 1 more galaxy has
contribution larger than 1 per cent, which is NGC7582 (6 per cent).
Regarding the FC component, contributions larger than 5
per cent were only found for two Seyfert 2s: NGC5135 and Mrk1210. Even
taking into account the ambiguity between the HII and FC components, it
is clear that FC fractions larger than 5--10 per cent are rare.

Considering the case when the fluxes are normalized to the flux
at $\lambda$3660\AA\ (Figure 1 right panel), we can see that the
percentage contribution from old metal rich stars is smaller,
while the contribution from younger stars, which are bluer, is
larger. The general trends seen for the spectra normalized at
$\lambda$5870\AA\ are also seen here, including the difference between
the AGN and the elliptical spectra. Also, the number of galaxies which
now show some contribution ($>1$ per cent) from $\leq 10$\ Myr stars or
from a FC increases to $\approx$50 per cent, or 25 per cent if we
consider only those galaxies where these components contribute with
more than 5 per cent.

The results for the elliptical galaxy IC4889 and the templates E1 and
E2 are very similar, so their average synthesis results are shown as
arrows in Figure 1.  Not surprisingly, the most noticeable
characteristic of their stellar populations is a dominant contribution
from old metal rich stars and almost no contribution from young stars.
The 10 Gyr and Z/Z$_{\odot}=0.6$\ component contributes with $>$60 per
cent to the light at 5870\AA, while the Z/Z$_{\odot}=0$\ component
contributes with 10--20 per cent. Stars with 10 Gyr and
Z/Z$_{\odot}<0$\ contribute with 5--10 per cent and 1 Gyr stars
contribute with 10--20 per cent, while the contribution from components
of 100 Myr and younger is smaller than 1 per cent. Note that these
results agree closely with those of Bica (1988).

\begin{figure*}
    \cidfig{22cm}{0}{140}{720}{720}{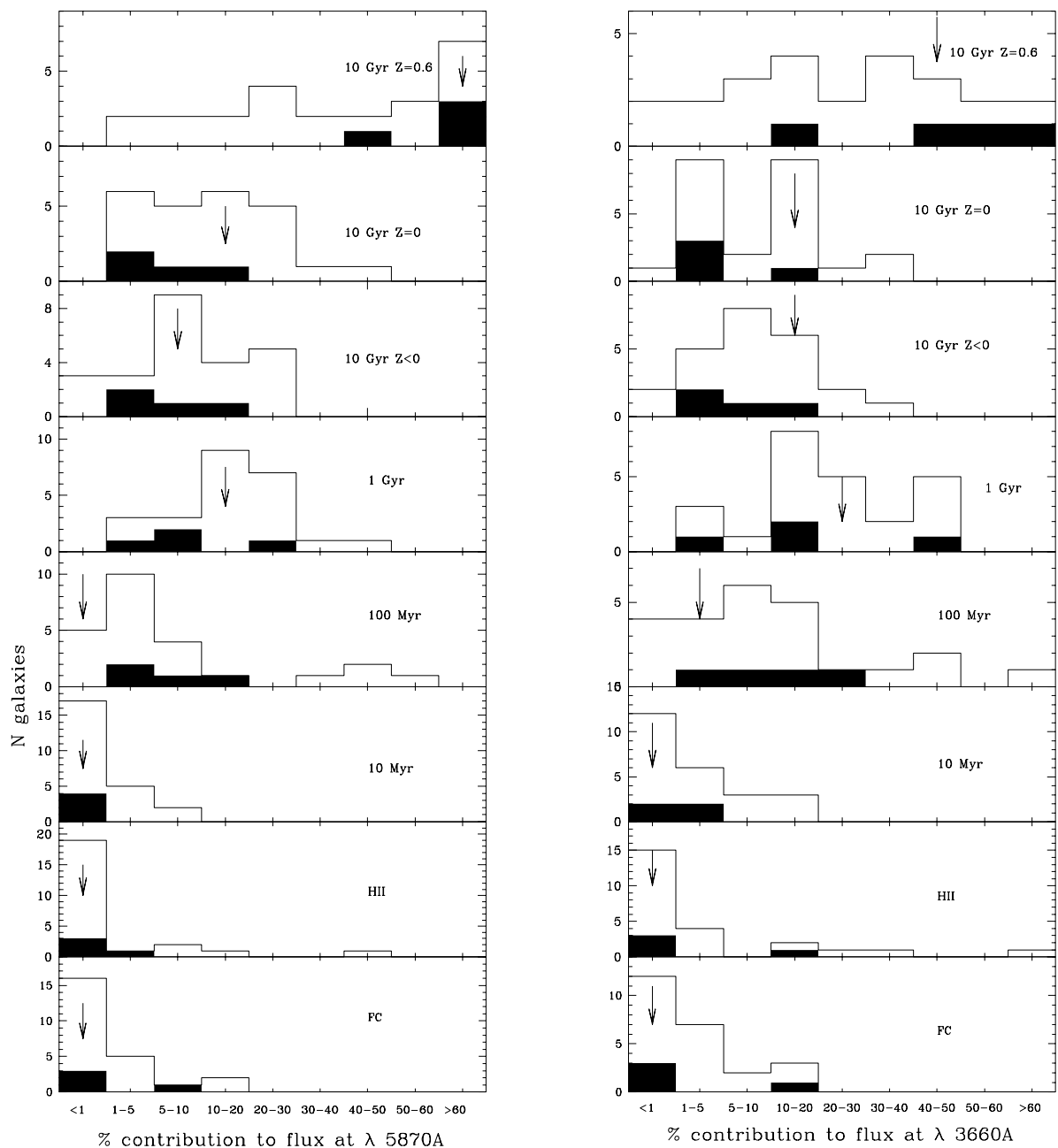}
    \caption{
Histogram of different age and metallicity components contribution,
normalized to
the light at $\lambda$5870\AA\ (left) and $\lambda$3660\AA\ (right).
The filled histograms represent the Radio Galaxies, the open ones
represent the cumulative histogram of Seyfert 2s and Radio Galaxies and
the arrows represent the Ellipticals. The panels show, from top to
bottom, the contribution from stars with 10 Gyr and
Z/Z$_{\odot}=0.6$\ to the light at $\lambda$5870\AA; stars with 10 Gyr
and Z/Z$_{\odot}=0.0$; the sum of stars with 10 Gyr and
Z/Z$_{\odot}=-1.0$\ and -2.0; the sum of the 3 metallicity bins of
stars with 1 Gyr; the sum of the 3 metallicity bins of stars with 100
Myr; the sum of the 3 metallicity bins of stars with 10 Myr; HII
region; and FC (F$_{\nu}\propto\nu^{-1.5}$).}
\end{figure*}

Comparing the stellar population synthesis results for Seyfert 2s with
those for elliptical galaxies, we can see that they differ
considerably in most cases, but can be meaningful for some galaxies.
The contribution of 10Gyr stars with  Z/Z$_{\odot} \ge
0$\ to the spectrum of Seyfert 2s is generally smaller than that
obtained for ellipticals. The contribution from low metallicity stars
with 10Gyr and, as well as stars with 1 Gyr to the spectrum of Seyfert
2's is usually equal or larger than in ellipticals, though there are
exceptions. The largest difference, however, occurs in the lower age
bins, particularly that corresponding to 100 Myr. Figure 1 shows that
whereas in ellipticals such stars contribute less than 1 per cent to
the flux at $\lambda$5870\AA, only 5 of the Seyfert 2s studied have
such a small contribution.  The contributions from stars with 10 Myr
and HII regions are also skewed towards values larger than those found
in ellipticals, but to a lesser extent than for 100 Myr stars.

The stellar population mixtures in 3 of the 4 Radio Galaxies is similar
to that of ellipticals, which is expected, since their host galaxies
are ellipticals. There is some difference in the contribution from
stars of 100 Myr, which is larger in Radio Galaxies than in
ellipticals, and in the contribution from stars with 1 Gyr and stars
with 10 Gyr and Z/Z$_{\odot}=0$, which is smaller than in ellipticals.
PKS0745-19 shows different results, with a smaller contribution of 10
Gyr, high Z/Z$_{\odot}$ stars and larger contributions of the younger
populations, which could be due to the fact that it is in the middle of
a cooling flow (Cardiel, Gorgas \& Arag\'on-Salamanca 1995).

\section{Discussion}

The results presented in the previous section are in sharp contrast
with the traditional view (e.g.\ Koski 1978) that the spectra of
Seyfert 2s are essentially composed of an old stellar population plus
an underlying FC. We have shown that: (1) Seyfert 2s present a wide
breadth of stellar population characteristics and thus cannot be
adequately represented by a single starlight template. (2) There are
substantial differences between the stellar populations of Seyfert 2s
and elliptical galaxies, particularly regarding the contribution
from stars with 10 Gyr and Z/Z$_{\odot}=0.6$ and stars with 100 Myr.
(3) The contribution of age $\le 10$\ Myr
stars and an FC is small, rarely exceeding 10 per cent of the light at
$\lambda$5870\AA.

Of particular interest in our analysis are the results for Mrk1210,
Mrk348, Mrk573, Mrk607, NGC1358 and 3C33. All these galaxies have been
previously studied by other authors, who have determined FC fractions
using the more traditional procedure of adopting the spectrum of an
elliptical galaxy (or the bulge of a normal spiral) as a starlight
template. According to Tran (1995a), for instance, the FC accounts for
25 per cent of the light at $\lambda$5500\AA\ in Mrk1210, while in
Mrk348 it contributes with 27 per cent. Our results for Mrk1210 show
that the FC contributes with only $\approx 10$ per cent of the light at
$\lambda$5870\AA, while stars with 10 Myr or younger, contribute with
11 per cent---note that these percentage contributions are
approximately constant over the 5500--5870 \AA\ interval. For Mrk348 we
find only $\approx 3$ per cent contribution from an FC and $\approx 2$
per cent from young stars. For 3C33, Koski (1978) found a 19 per cent
FC contribution to the light at $\lambda$5500\AA, while the synthesis
yields less than $1$ per cent from either FC or young stars. For
Mrk573, Mrk607 and NGC1358, Kay (1994) estimated an FC fraction at
$\lambda$4400\AA\ of 20 per cent, 10 per cent and 17 per cent,
respectively. Our results show only a 1 per cent FC for  Mrk573 and
even less for NGC1358 and Mrk607, while young stars contribute with
$\approx$2 per cent to the spectrum of Mrk573 at $\lambda$4510\AA\ and
$<1$ per cent for NGC1358 and Mrk607.

In order to check if the smaller FC contribution found by us is real,
or if it is just due to the spectral synthesis technique used, we
repeated the spectral synthesis allowing only FC contributions, at
$\lambda$5870\AA, larger than 25 per cent for Mrk1210 and Mrk348, 20
per cent for 3C33 and Mrk573, 15 per cent for NGC1358 and 10 per cent
for Mrk607.  These values are close to the ones found by other authors
in the literature.  With the exception of Mrk1210, it was not possible
to find combinations that would fit the observed spectra for the
windows previously used. 

We have also repeated this test for all galaxies in the sample,
allowing only a minimum FC or HII region contribution of 5 per cent
at $\lambda$5870\AA. Again, for most of the galaxies, the results of
this forced synthesis yielded poorer $\chi^2$ than that of using all
possible contributions of the FC or the HII
region, indicating that these components contribute with less than 5
per cent to the spectrum. The exceptions are the galaxies with
circumnuclear starburst, for which the results obtained forcing a
larger contribution from these components gave a better $\chi^2$.
Due to the ambiguity of the HII regions and FC component,
they also show a better fit when we force a larger contribution of this
component.

The above discrepancy between our results and those obtained by other
authors is exacerbated by the fact that our spectra were obtained with
an aperture 2--4 times smaller than those used by the above authors,
which should lead to even larger FC fractions, contrary to what is
seen. On the other hand, the synthesis results for Mrk348, Mrk1210 and
NGC1358 are in good agreement with those obtained in Paper~II,  where a
spectral decomposition using an extranuclear extraction as a stellar
population template for the nucleus was carried out, leading to the
conclusion that the FC contribution was smaller than 10 per cent for
these 3 galaxies at $\lambda$5870\AA. Furthermore, the small
contributions from a FC or young stars found in the synthesis fit
nicely with the main result of Paper~I, namely, that for most of these
galaxies the nuclear Ws are similar to the extranuclear ones,
indicating no dilution of the nuclear absorption features, which would
be expected if a FC or a blue stellar population were present at the
nucleus.  The results presented here are thus not isolated, but
entirely consistent with the previous findings of Papers~I and II.
Taken together, these results suggest that previous measurements of the
FC contribution to the nuclear spectrum of Seyfert 2s have been
overestimated due to the use of an inadequate stellar population
template.

In summary, our spectral synthesis analysis showed that the usual
assumption that the nuclear stellar population of Seyfert 2 galaxies
can be represented by an elliptical galaxy template is not necessarily
correct.  With the exception of the Radio Galaxies, most of the Seyfert
2s do not have as large a contribution from old metal rich stars as
ellipticals do. Elliptical galaxies have deep absorption lines (large
Ws), which, when compared with the spectrum of Seyfert 2s, create the
impression that these lines are diluted by a large proportion of an
FC.  What actually happens is that Seyfert 2s and ellipticals have
different stellar populations.

The main difference obtained for 19 of the 24 galaxies, when normalized
at $\lambda$5870\AA, or 16 of 24 when normalized at $\lambda$3660\AA,
is the larger  contribution of ``intermediate age'' stars of $\approx
100$\ Myr to the spectrum of Seyfert 2s and Radio galaxies, compared to
ellipticals. This result points to some sort of connection between
star-formation and nuclear activity, which we speculate can be related
to the process of fueling the AGN. Several models explain the fueling
of the nuclear engine in terms of galaxy interactions or mergers (Byrd
\et 1986, 1987; Lin, Pringle \& Rees 1988; Hernquist \& Mihos 1995).
According to these models, the interaction between two galaxies brings
the gas to the nuclear region (inner hundred parsecs), where it is
shocked and compressed, producing a starburst and feeding the nucleus.
The starburst phase starts 100--300 Myr prior to the feeding of the
nuclear engine and can spread throughout this period.  This time scale
agrees with the larger contribution from 100 Myr stars found in the
synthesis, as well as with the contribution from younger populations
found in some of the galaxies.

Finally, we would like to discuss the implications of our results to
the nature of FC2. We conclude that this component is due, in most of
the cases, to a template mismatch, as most of the previous works used
elliptical templates to subtract the stellar population contribution
from the nuclear spectrum of AGN, overestimating the amount of FC.  As
shown here, the role of FC2 can be played by stars with 100 Myr.
Elliptical galaxies have $<1$ per cent contribution from 100 Myr stars,
while for 19 of the 24 AGN's studied here, the contribution from these
stars to the light at $\lambda$5870\AA\ is larger than 1 per cent,
or 9 of 24 galaxies if we consider contributions larger than 5 percent.
In a few cases the FC2 can also be due to stars with age $\leq 10$ Myr,
which also contribute with less than 1 percent to the light at
$\lambda$5870\AA\ in ellipticals.
In 4 of the 24 galaxies, stars with $<10$ Myr contribute with $>5$ per cent,
or 9 of 24 if we consider galaxies with contribution $>1$ per cent.

{\bf Acknowledgements:}

This work was partially financed by CNPq, FINEP, FAPERGS and
FAPEU-UFSC.

\end{document}